\documentclass[journal]{IEEEtran}
\usepackage{cite}
\usepackage{amsmath,amssymb,amsfonts}
\usepackage{graphicx}
\usepackage{algorithm}
\usepackage[noend]{algpseudocode}
\usepackage{pifont}
\usepackage{textcomp}
\usepackage{booktabs,multirow}
\usepackage[utf8]{inputenc}
\usepackage{url}
\usepackage{hyperref}

\def\BibTeX{{\rm B\kern-.05em{\sc i\kern-.025em b}\kern-.08em
    T\kern-.1667em\lower.7ex\hbox{E}\kern-.125emX}}
\begin{document}
\title{TEPI: \underline{T}axonomy-aware \underline{E}mbedding and \underline{P}seudo-\underline{I}maging for Scarcely-labeled Zero-shot Genome Classification}
\author{Sathyanarayanan N. Aakur, \IEEEmembership{Member, IEEE}, Vishalini R. Laguduva, and Priyadharsini Ramamurthy, and Akhilesh Ramachandran
\thanks{
Submitted on 08 June, 2023. Revised on 27 September, 2023. 
This research was supported in part by the US Department of Agriculture (USDA) grants AP20VSD and B000C011.
}
\thanks{Sathyanarayanan N. Aakur and Vishalini R. Laguduva are with the Department of Computer Science and Software Engineering, Auburn University, Auburn, AL 36849 (e-mail: \{san0028,vlr0013\}@auburn.edu. Work was partially done while they were at OSU.}
\thanks{Priyadharsini Ramamurthy is with the Department of Computer Science at Oklahoma State University, Stillwater, OK 74078 USA.
}
\thanks{Akhilesh Ramachandran is with the Oklahoma Animal Disease Diagnostics Laboratory, College of Veterinary Medicine, Oklahoma State University, Stillwater, OK 74078 (e-mail: rakhile@okstate.edu)}
\thanks{Corresponding authors: SN Aakur, A. Ramachandran}}

\maketitle
\begin{abstract}
A species' genetic code or genome encodes valuable evolutionary, biological, and phylogenetic information that aids in species recognition, taxonomic classification, and understanding genetic predispositions like drug resistance and virulence. However, the vast number of potential species poses significant challenges in developing a general-purpose whole genome classification tool. Traditional bioinformatics tools have made notable progress but lack scalability and are computationally expensive. Machine learning-based frameworks show promise but must address the issue of large classification vocabularies with long-tail distributions. In this study, we propose addressing this problem through zero-shot learning using TEPI, \underline{T}axonomy-aware \underline{E}mbedding and \underline{P}seudo-\underline{I}maging. We represent each genome as pseudo-images and map them to a taxonomy-aware embedding space for reasoning and classification. This embedding space captures compositional and phylogenetic relationships of species, enabling predictions in extensive search spaces. We evaluate TEPI using two rigorous zero-shot settings and demonstrate its generalization capabilities qualitatively on curated, large-scale, publicly sourced data. 
\end{abstract}

\section{Introduction}\label{sec:intro}
The genome sequence of an organism offers valuable insights into disease susceptibility (e.g., cancer in humans and animals) and quantitative traits like milk production in animals~\cite{cockett2008genome} or crop yield in plants~\cite{brinton2020haplotype}. Recent technological advancements have made nucleic acid sequencing (DNA or RNA sequencing) of animals, plants, and microbes affordable and commonplace~\cite{criscuolo2019fast}. In microbiology, whole and partial genome sequencing is frequently used to study the origins, pathogenicity, and phylogenetic relationships of bacterial and viral organisms~\cite{koser2012routine}. Genome sequencing is crucial for phylogenetic classification and determining virulence potential, such as toxin production, antibiotic resistance, or invasive abilities, in bacterial organisms~\cite{razin1998molecular}.
\begin{figure}[t]
    \centering
    \includegraphics[width=0.48\textwidth]{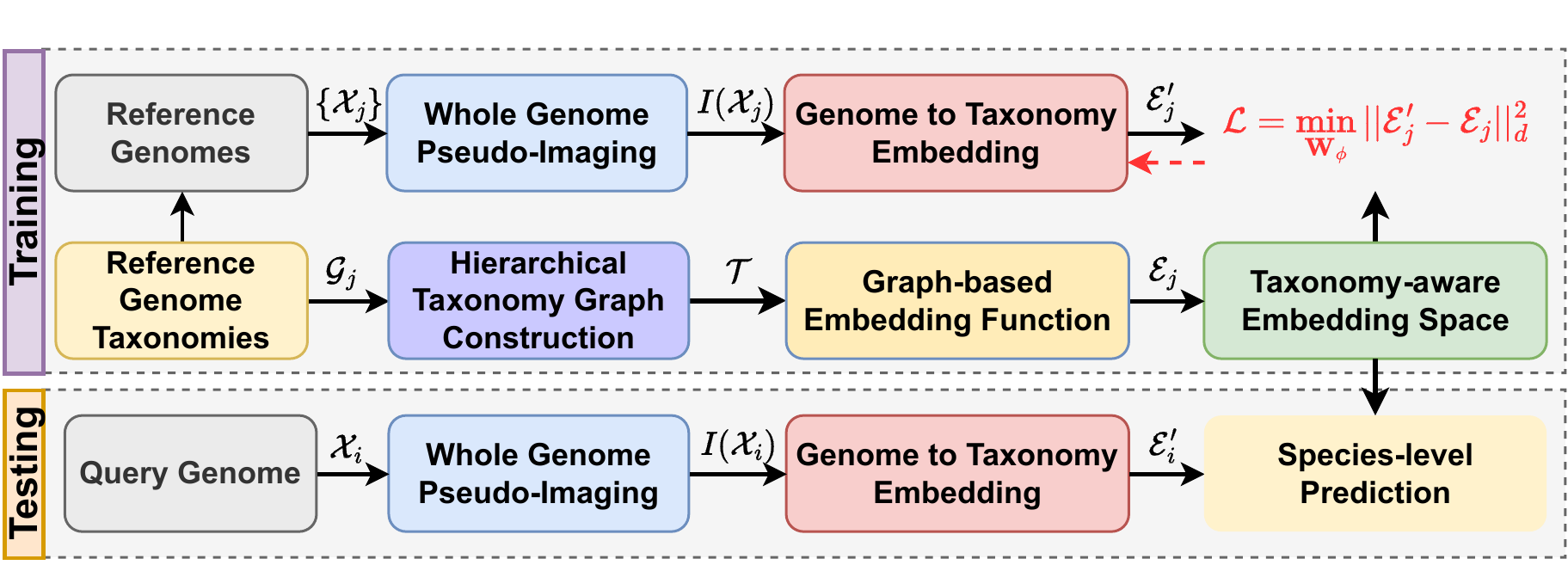}
    \caption{Illustration of the \textbf{Overall Workflow} of the proposed framework. We learn a taxonomy-aware embedding space during training that is used to predict taxonomies of query genomes during testing. Dotted red lines indicate learning. }
    \label{fig:workflow}
\end{figure}

However, nucleic acid sequencing generates large amounts of microbial genome data. Traditional bioinformatic pipelines for genome analysis, particularly for whole-genome-based phylogenetic studies, are computationally intensive and time-consuming~\cite{criscuolo2019fast,ye2006blast}. Developing efficient computational frameworks for processing whole genomes is essential to enable epidemiological, diagnostic, source attribution, and phylogenetic applications. Machine learning approaches, driven by advances in deep learning, offer promising avenues for genomic data analysis. However, successful machine learning applications face two key challenges. 
Machine learning approaches, particularly those based on deep learning, have been successful in handling classification problems where the label space is in the order of thousands (under an \textit{extreme} classification problem setup~\cite{liu2017deep}). They require thousands of examples per class to achieve reasonable performance. 
Genome classification, however, involves an enormous number of classes, with estimates placing the number of bacterial species between $10^7$ and $10^9$~\cite{curtis2002estimating}. Hence, scaling machine learning to such a large label space requires robust representation learning~\cite{bengio2013representation}. 
Additionally, the distribution of labeled data is highly imbalanced, i.e., not all species are equally represented in the training and test data. This imbalance poses a critical challenge to capturing the entire spectrum of species and hence scaling deep learning algorithms to this important problem.  

\begin{figure*}[t]
\centering
    \includegraphics[width=\textwidth]{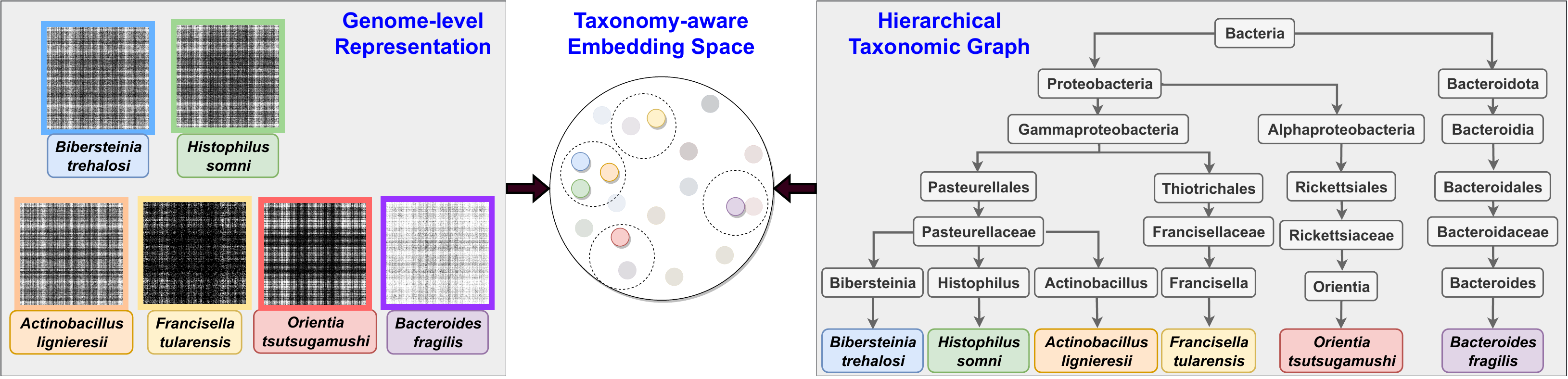}
    \caption{\textbf{Overall architecture} of the proposed TEPI model. We aim to learn a common, taxonomy-aware representation space to which each species' genome representations can be mapped to ensure generalizable, zero-shot classification performance.}
    \label{fig:arch}
\end{figure*}

In this study, we address these challenges through zero-shot learning with limited labeled data. 
Zero-shot learning~\cite{xian2017zero} aims to recognize classes that may not have been seen during training. 
Given the large number of species, we adopt a deductive reasoning approach by mapping a species' whole genome to a learned, compositional representation within a predefined search space for accurate identification. Unlike brute-force search-based algorithms, our scalable framework utilizes zero-shot learning to create taxonomy-aware, generalizable genome representations. Figure~\ref{fig:arch} presents an overview of our approach. During training, we learn a mapping function that projects whole genomes (represented as ``pseudo-images'') into a compositional, hierarchical embedding space. Genome classification involves traversing the taxonomy-aware embedding space to find the best match during inference. This zero-shot mechanism paves the way to build general-purpose diagnostic applications that can generalize beyond the species seen during training and move beyond targeted pathogen recognition~\cite{aakur2021metagenome2vec,aakur2021mg}. 

A preliminary version of this work appears as MG-NET~\cite{aakur2021mg}, where we proposed a novel representation mechanism, called a \textit{pseudo-image}, for genome sequences. Building upon the success of visual computing techniques~\cite{bengio2013representation,xian2017zero}, we structured genomes as images using a co-occurrence-based formulation. This paper significantly extends the original pseudo-imaging approach. Firstly, we expand the representation to encompass entire genomes instead of single reads from metagenomes, which are a collection of DNA sequences extracted from clinical or environmental samples and can be composed of nucleic acid from different organisms. Representing whole genomes is challenging since it requires capturing different phylogenetic properties of the organism in an expressive representation. 
Secondly, we introduce zero-shot learning for the taxonomic classification of bacterial genomes through a hierarchical, taxonomy-aware embedding space. Lastly, we extensively analyze the representation learning capability of pseudo-imaging-based frameworks for extreme multiclass classification.

The proposed TEPI framework offers three significant \textbf{contributions}: (i) we are among the first to address the challenge of zero-shot genome recognition, (ii) we demonstrate that constructing a taxonomy-aware, hierarchical embedding space enables effective generalization to unseen classes with limited labeled training data, and (iii) through a large-scale whole-genome study, we show that the proposed TEPI framework can learn compositional representations and generate highly relevant predictions.
This paper is structured as follows. Section~\ref{sec:related} reviews prior relevant works and their relation to our proposed techniques. Section~\ref{sec:proposed} presents the TEPI framework, providing the necessary background information and mathematical notations. The experimental setup, evaluation metrics, baselines, and results are outlined in Section~\ref{sec:results}, along with a discussion on its limitations and the opportunities in zero-shot genome analysis.
Finally, in Section~\ref{sec:conclusion}, we conclude with a brief discussion of the approach and future research directions.

%
\begin{figure*}[t]
    \centering
    \includegraphics[width=0.99\textwidth]{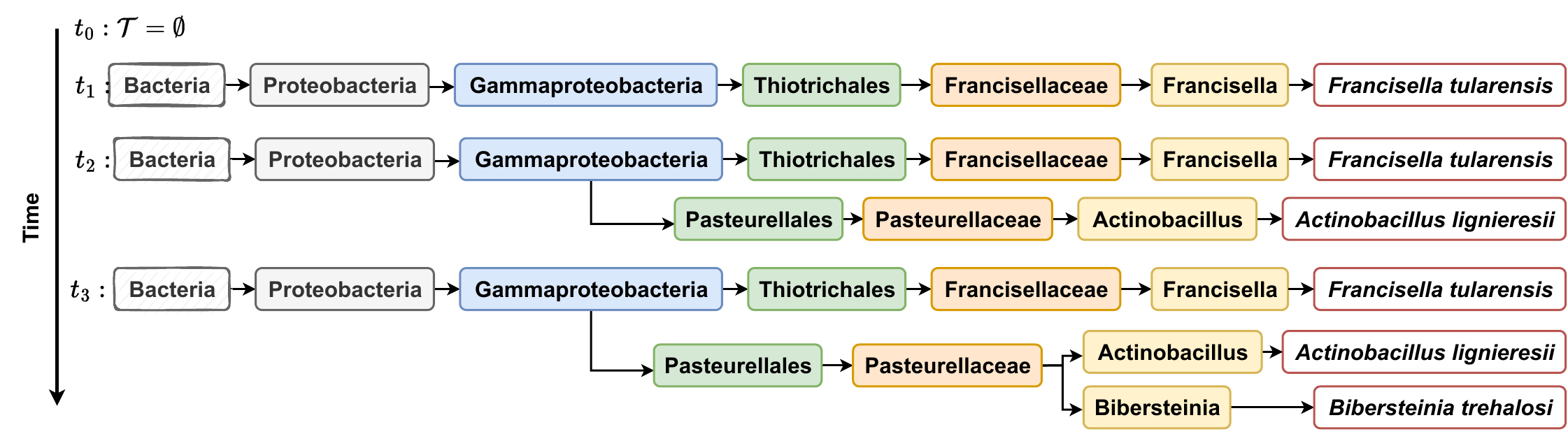}
    \caption{Iterative construction of the hierarchical taxonomy graph to capture phylogenetic relationships between species. We iteratively add nodes based on the species taxonomy from an empty graph to capture the inherently compositional relationships. }
    \label{fig:building_graph}
\end{figure*}
\section{Related Work}\label{sec:related}
\textbf{Traditional bioinformatics platforms} such as Kraken~\cite{wood2014kraken,wood2019improved}, BLAST~\cite{ye2006blast}, Mauve~\cite{darling2004mauve}, and Centrifuge~\cite{kim2016centrifuge} have been widely used for genome analysis. These algorithms involve matching nucleotide sequences to a predetermined search space and calculating the statistical significance of the matches to infer phylogenetic relationships between species. For instance, BLAST utilizes a greedy algorithm to identify the best alignment between query DNA sequences and those in the search space. The Kraken series~\cite{wood2014kraken,wood2019improved} employs a lightweight algorithm to align k-mers against a pathogen database. Centrifuge~\cite{kim2016centrifuge} utilizes metagenome-specific indexing mechanisms for fast and accurate classification. However, these approaches are computationally expensive due to their reliance on costly alignment steps and brute-force matching. While recent advancements have improved inference speed~\cite{criscuolo2019fast}, scalability remains an ongoing challenge. Furthermore, these methods are limited by the user-defined search space, which determines the best match based on alignment.

\textbf{Machine learning} approaches have provided valuable opportunities for analyzing DNA sequences obtained directly from environmental samples~\cite{ching2018opportunities}. Deep learning models, in particular, have been extensively investigated for representation learning in metagenome sequences, addressing various tasks such as capturing nucleotide representations using reverse complement convolutional neural networks (CNNs) and Long Short Term Memory networks (LSTMs)~\cite{bartoszewicz2020deepac}, predicting metagenome sequence taxonomy with depth-wise separable convolutions~\cite{busia2019deep}, predicting genomic sub-compartments and disease genes using graph representations~\cite{ashoor2020graph,hwang2019humannet}, learning representations with bidirectional LSTMs and \emph{k}-mer embeddings with self-attention for sequence taxonomy prediction~\cite{liang2020deepmicrobes}, and employing ResNet to learn metagenome representations for taxonomy prediction~\cite{He2016CVPR}. 
Supervised learning methods like MGNet~\cite{aakur2021mg}, GraDL~\cite{narayanan2020genome}, and MG2V~\cite{aakur2021metagenome2vec} have been proposed for sequence-level identification tasks. Other approaches, such as Metagenome2Vec~\cite{queyrel2021towards}, focus on disease prediction directly from metagenome sequences. However, these methods are trained in a supervised manner and are designed to generalize to unseen species without specifically addressing the challenges associated with phylogenetic classification using whole genome sequences.

\textbf{Pseudo-imaging} is a mechanism of representing data from other sensing modalities, such as ultrasound~\cite{sun2018research}, as images to extract richer contextual information that is not captured in conventional imaging. It has been extensively employed in astrophysics~\cite{nelson2006pseudo} and medicine~\cite{pennec2003tracking}, to name a few. In medicine, pseudo-CT estimations have been extracted from other sensors such as MRI imaging~\cite{leu2020generation} and ultrasound deformation fields~\cite{sun2018research}. In astrophysics, spinning prism-based sensors and signal processing techniques encode star-specific events such as explosions and are represented as images to extract scene-specific information about the event and its location~\cite{nelson2006pseudo}. Such representations of unstructured data as images allow us to leverage advances in deep learning models, particularly convolutional neural networks (CNNs)~\cite{He2016CVPR,simonyan2014very}, for extracting robust, discriminative features from pseudo-images. 
Pseudo-images have yet to be explored as extensively in genome analysis, although their application holds much promise in extracting structural properties from sequential data. Some works have used the idea of alternative representations using images for \textit{metagenome}-based analysis for disease classification and host phenotype prediction. For example, Self-Organizing Maps (SOM)~\cite{nguyen2019metagenome,nguyen2020growing} have been proposed to represent metagenomes as images, which are then processed using CNNs for disease classification. 
A matrix-based phylogenetic tree representation~\cite{reiman2020popphy} was used as a metagenome representation mechanism for predicting host phenotype using CNNs. However, representing the whole genome as pseudo-images for species-level recognition and similarity analysis has not been explored in prior literature. 

\textbf{Zero-shot learning}, which has been extensively explored in the computer vision~\cite{karessli2017gaze,wu2020multi} and text processing communities~\cite{zhang2019integrating,merrillees2021stratified}, aims to extend supervised machine learning to handle novel, unseen classes. In this paradigm, a semantic vector representation of target classes is learned, and inputs (e.g., images or text) are mapped to this space for classification. However, traditional zero-shot learning approaches assume the availability of abundant annotated training data for the \textit{seen} classes. This assumption does not hold for genome sequences due to a significant long-tail distribution problem and an extremely large number of target classes.

This work addresses the challenge of zero-shot learning in genome classification, specifically in a sparsely labeled setting. We adopt the traditional zero-shot learning pipeline but propose leveraging compositional relationships derived from well-defined taxonomies to learn a hierarchical and generalizable embedding space. Genome sequences can then be mapped to this embedding space, enabling classification even for unseen classes.

\begin{figure*}[t]
    \centering
    \begin{tabular}{cccc}
    \includegraphics[width=0.22\textwidth]{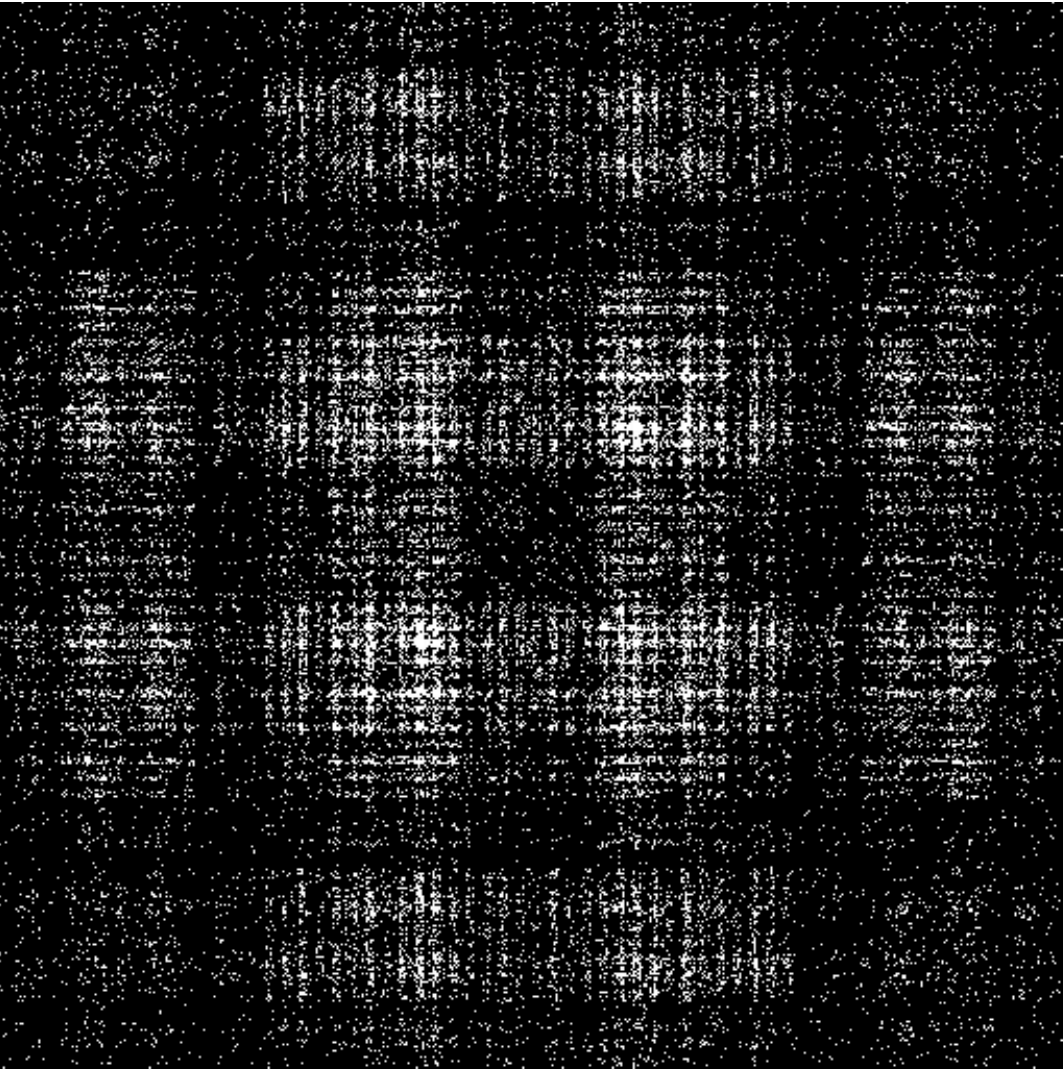}     &  
    \includegraphics[width=0.22\textwidth]{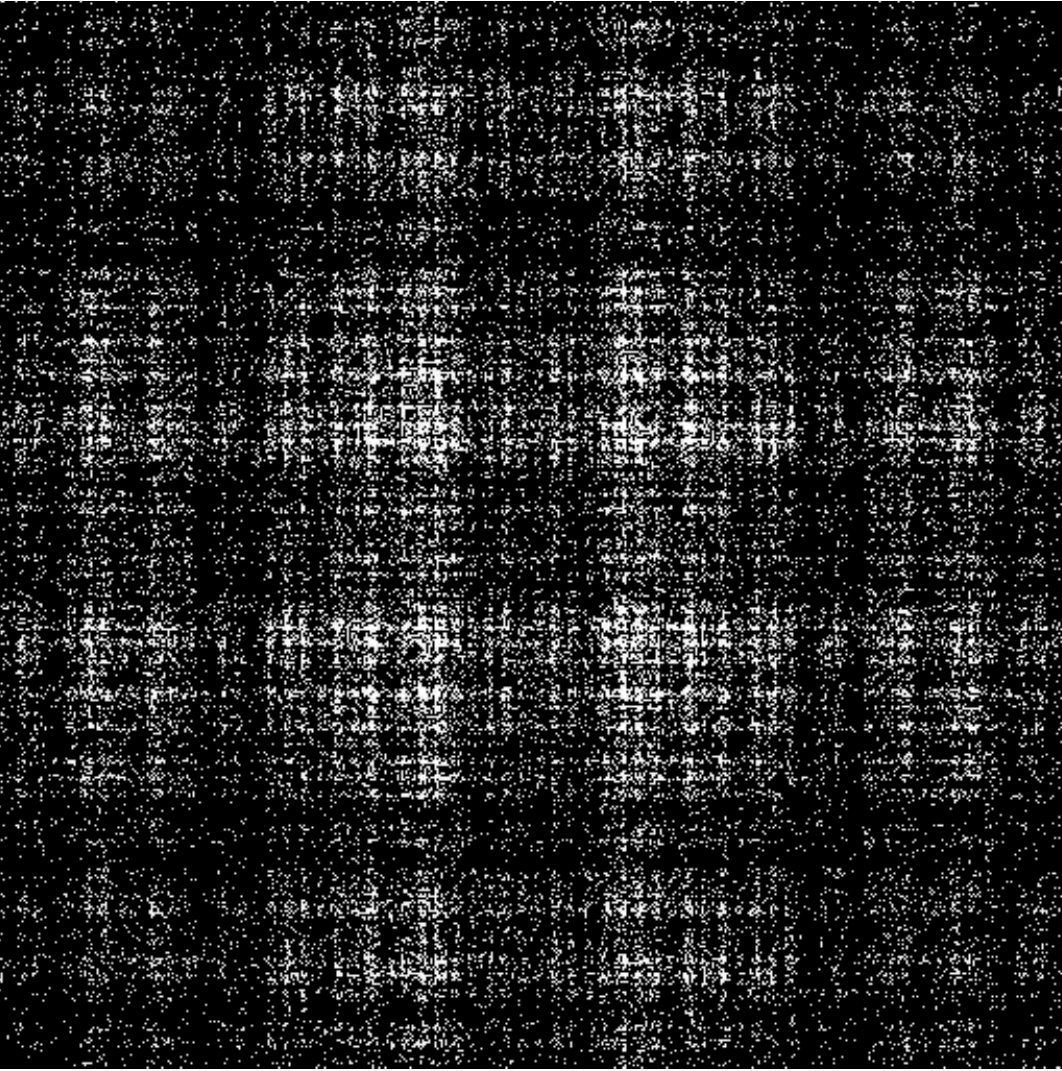}     &
    \includegraphics[width=0.22\textwidth]{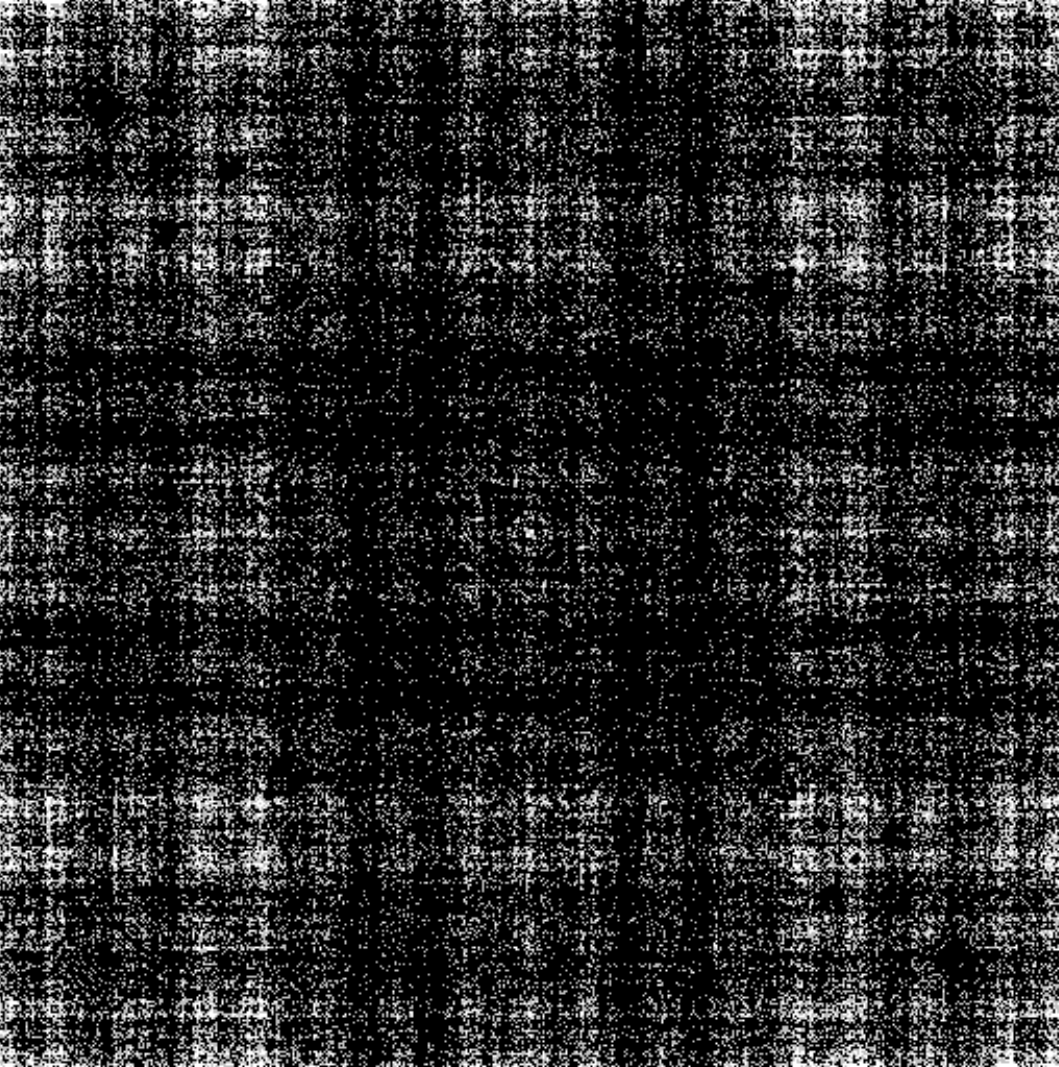}     &  
    \includegraphics[width=0.22\textwidth]{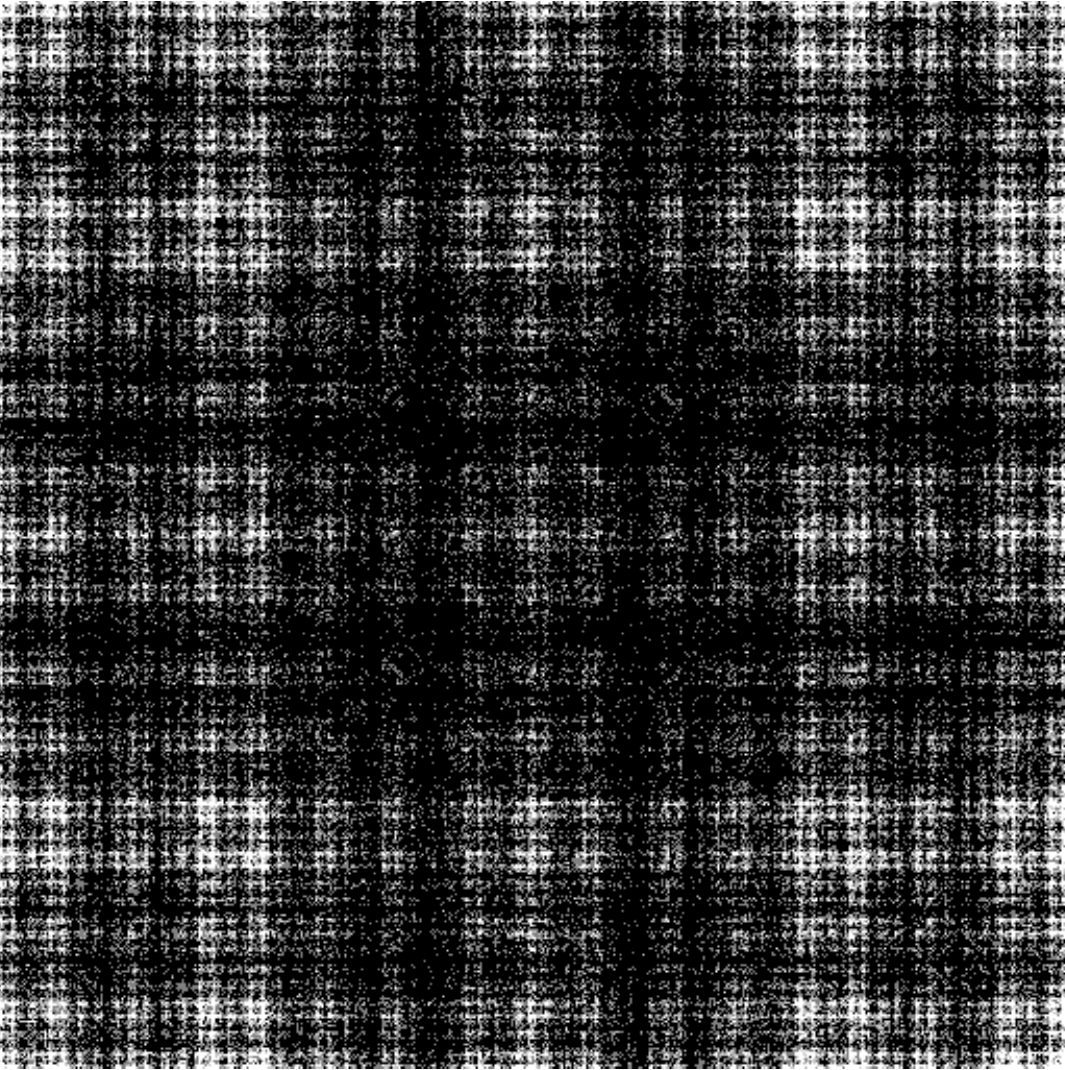}     \\
    (a) & (b) & (c) & (d)\\
    \end{tabular}
    
    \caption{\textbf{Genome-level complement-aware, pseudo-images} of genomes from (a) \textit{Mycobacterium avium}, (b) \textit{Mycobacterium tuberculosis}, (c) \textit{Francisella tularensis} and (d) \textit{Clostridium botulinum}. Note the similar patterns between the former two images, which belong to species from the same genus \textit{Mycobacterium}. }
    \label{fig:PI_Viz}
\end{figure*}

\section{Proposed Framework: TEPI}\label{sec:proposed}
\textbf{Problem statement.}
This work addresses the problem of scarcely-labeled zero-shot learning for genome classification. The objective is to learn a generalizable embedding $\mathcal{E}_i$ for a given whole genome sequence $\mathcal{G}_i = \{b_1, b_2, b_3, \ldots b_m\}$ of a species $s_i \in \mathcal{S}$, where $b_i\in \{A, T, C, G\}$ represents a nucleotide base. Each species can be characterized by its taxonomic lineage, which represents the hierarchical description of the organism based on shared characteristics with other organisms. The taxonomy assigns organisms to different levels, including life, domain, kingdom, phylum, class, order, family, genus, species, and optionally sub-species, reflecting the phylogenetic relationships between organisms. The aim is to learn a vector representation that captures this hierarchy and develop a mechanism to map whole genomes to this vector space, enabling species-level recognition with very few labeled examples per species (${<=}5$).

\textbf{Overview.}
The overall goal of the proposed approach (TEPI), illustrated in Figure~\ref{fig:workflow}, is to learn robust features from genome data that allow for efficient taxonomic classification with a limited number of training examples. We propose a genome-to-taxonomy mapping function that facilitates species-level identification from whole genome sequences. Efficient representations are formulated for both the genome and the taxonomy to enable the learning of this mapping. Specifically, complementary-aware pseudo-imaging is used to represent the genome, and a taxonomy-aware embedding space is learned to capture the compositional relationships between species based on their taxonomic lineage. A detailed description of the proposed approach is provided below.

\subsection{Learning a Taxonomy-aware Embedding Space}\label{sec:embed}
\begin{figure*}[t]
    \centering
         \includegraphics[width=0.85\textwidth]{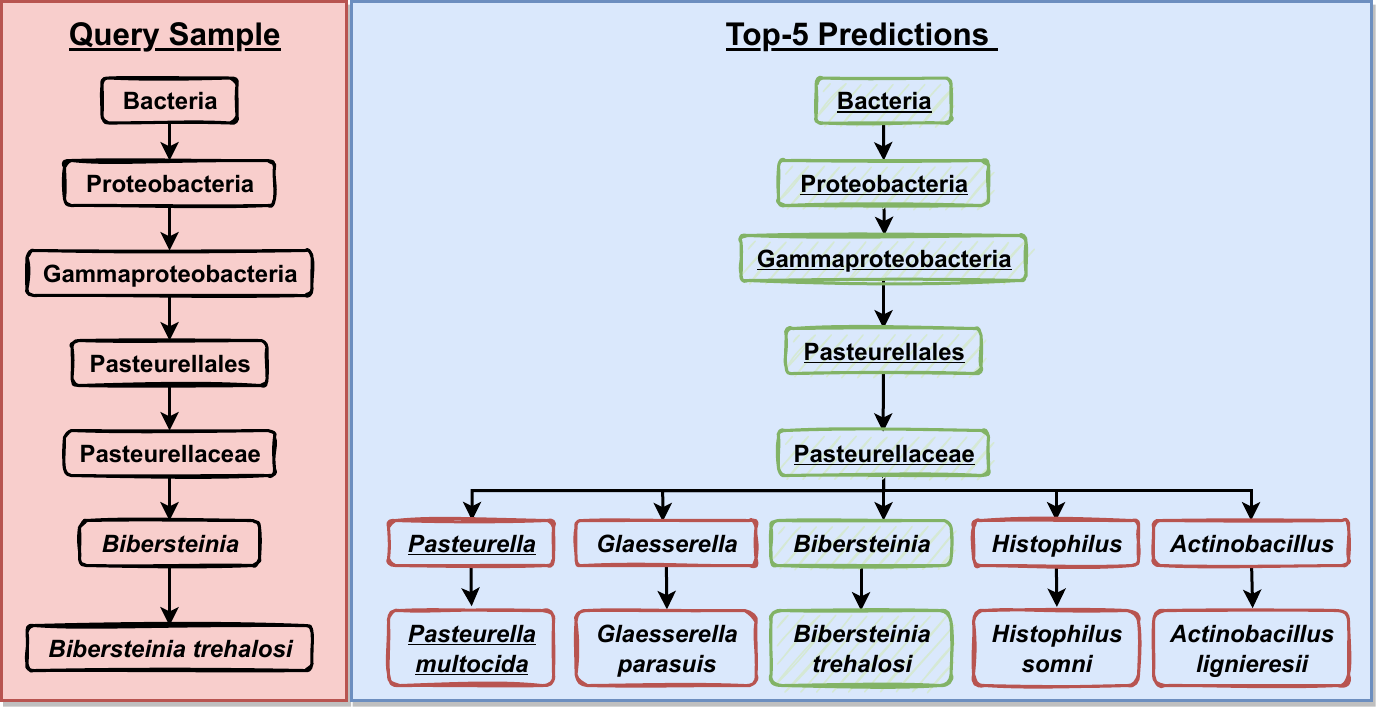}  \\
    \caption{\textbf{Qualitative visualization} of the predicted taxonomy tree from top-5 predictions (on the right) for a given query genome (left) for an unseen \textit{genus}. All retrieved species belong to the same family, indicating the compositional structure of the learned embedding space. The top-1 prediction is underlined, and the correct ones are in green. }
    \label{fig:qual_genus}
\end{figure*}

TEPI is centered around a taxonomy-aware embedding space that can be used for species recognition based on genome sequences. This embedding space should capture the phylogenetic properties of each species by encoding the hierarchical relationships established by their taxonomic lineage. The higher the similarity in the embedding space between two genomes, the more closely related they are in the taxonomy. Capturing this property in the embedding space is essential for effectively recognizing species outside the training data. 
To achieve this, we construct a taxonomic graph $\mathcal{T} = \{{V}, {E}\}$, where ${V}$ represents the nodes in the taxonomy at different levels of the hierarchy, connected by edges ${E}$. The construction of this taxonomic graph $\mathcal{T}$ is done recursively in a top-down manner, starting from the \textit{kingdom} level and progressing to the \textit{species} level, and optionally the \textit{sub-species} level if applicable. Figure~\ref{fig:building_graph} provides an example of how such a graph can be constructed. Traversing this graph from one species node to another using a random walk reveals the phylogenetic similarity between the two species. The length of the walk reflects their similarity, while the path taken provides information on their relationship. For instance, in Figure~\ref{fig:building_graph}, we can infer that the representations for \textit{Bibersteinia trehalosi} and \textit{Actinobacillus lignieresii} should be closer together than to \textit{Francisella tularensis} since the former two belong to the same family (\textit{Pasteurellaceae}) and can be reached within three steps. However, all three species belong to the same class \textit{Gammaproteobacteria}, and hence, they can be reached within eight steps. Figure~\ref{fig:qual_genus} demonstrates this property, where the proposed embedding successfully guides taxonomy navigation and provides highly relevant predictions for a genome whose corresponding genus (\textit{Bibersteinia}) and species (\textit{Bibersteinia trehalosi}) were not observed during training. Section~\ref{sec:qual} provides additional discussion on this phenomenon.
With the constructed taxonomic graph and its inherent compositional properties, we employ the \textit{node2vec} algorithm~\cite{grover2016node2vec} to learn a low-rank embedding space. \textit{node2vec} offers a flexible objective that preserves the neighborhood structure for each node and captures it in the embedding space using simulated biased random walks. This formulation captures the phylogenetic relationships between species as described above. By incorporating hyperparameters such as the return and in-out parameters, \textit{node2vec} combines depth-first search (DFS)-like and breadth-first search (BFS)-like neighborhood exploration and employs negative sampling to optimize a Skip-gram model~\cite{mikolov2013distributed}. After optimization, we obtain a set of embeddings $\mathcal{E}_i$ for each target species $s_i \in \mathcal{S}$. These embeddings provide taxonomy-aware, hierarchy-based representations for each species.

\begin{algorithm}[t]
\caption{Constructing a pseudo-image from a whole genome sample for a given k-mer size $k$.}\label{alg:cap}
\begin{algorithmic}
\Require Sequence reads $X_i = \{X_0, X_1, \ldots X_N\}, \forall X_i \in \mathcal{X}_i$
\Ensure Pseudo-image $I_r$
\Procedure{relativeCoOccurrence}{$x_i, x_j$, $I_r$}
\State $e_{i,j} \gets I_r[i,j]$ \Comment{Current co-occurrence frequency}
\State $e^\prime_{i,j} \gets e_{i,j} + 1$ \Comment{New co-occurrence frequency}
\State $r \gets \frac{e_{i,j}}{(\Vert e_{i,j} - e^{\prime}_{i,j} \Vert_2)}$ \Comment{Update co-occurrence}
\State $e_{i,j} = 2 \sqrt{max(r - 1, 1)} + (min(r-2,2) + 2)$ 
\State \textbf{return} $e_{i,j}$\Comment{Relative co-occurrence frequency}
\EndProcedure
\Procedure{NormalizeImage}{$I_r, \lambda_m$}
\State $N\gets\sum_{i=0}^{4^k}{\sum_{j=0}^{4^k}{I[i,j]}}$
\For{$i \in 4^k$}
\For{$j \in 4^k$}
\If{$I_r[i,j] > \lambda_m$}
\State$I_r[i,j]\gets 255*I[i,j]/N$
\Else
\State$I_r[i,j]\gets 0$
\EndIf
\EndFor
\EndFor
\State \textbf{return} $I_r$\Comment{Normalized Pseudo-Image}
\EndProcedure
\State $I_r \gets ones(4^k,4^k)$ \Comment{Initial pseudo-image}
\For{$ X_i \in \mathcal{X}_i $} \Comment{Iterate through each sequence}
\For{$ x_i, x_j \in X_i $} \Comment{Iterate through successive k-mers}
\State $I_r[i,j]\gets$\Call{relativeCoOccurrence}{$x_i, x_j$, $I_r$}
\EndFor
\EndFor
\State $I_r\gets$ \Call{NormalizeImage}{$I_r, \lambda_m$}
\State \textbf{return} $I_r$\Comment{Final pseudo-image}
\end{algorithmic}
\end{algorithm}
\subsection{Representing Genomes as Pseudo-Images}\label{sec:PI}
The other key component of our framework involves finding a genome representation that captures inter- and intra-species variations at the nucleotide read level. We build upon the pseudo-imaging concept introduced in our previous work, MGNet~\cite{aakur2021mg}, to achieve this. We devise a pseudo-imaging technique to represent genome sequences as images to capture pathogen-specific \textit{fingerprints}.
To create the pseudo-image of a whole genome sequence, we capture the relative co-occurrence between \textit{k-mers}, or smaller sub-sequences of each genome, in a histogram-based representation and create a pseudo-image, following the success of Gray Level Co-occurrence Matrices (GLCM)~\cite{ballerini2009query,bagari2018combined}. Figure~\ref{fig:PI_Viz} visualizes some examples of generated pseudo-images. The intensity of each pixel represents the relative frequency with which two k-mers (or nucleotide subsequences) co-occur in a genome, with the resulting image forming a species-specific pattern that captures the taxonomic similarity between species. We describe the process of building pseudo-images from genomes in detail below. 

We consider each whole genome ($\mathcal{X}_i$) to be a collection of nucleotides drawn from a set of bases $\{A, T, C, G\}$, sequenced as ``reads'' ($X_0, X_1, \ldots, X_n \in \mathcal{X}_i$). Each sequence read is broken down into smaller atomic sub-sequences of length $k$, called \textit{k-mers}, using a sliding window of length $k$ and stride $s$. We denote each k-mer as $x_j$, where $x_0, x_1, \ldots, x_l \in X_i$. We compute the \textit{relative} co-occurrence between two k-mers ($x_i$ and $x_j$) by defining a function $f_r(x_i, x_j)$ that iteratively updates their relative co-occurrence using their observed co-occurrence across all sequence reads in the whole genome sequence $\mathcal{X}_i$. Formally, we define this function as follows.
\begin{equation}
    f_r(x_i, x_j) {=} \sigma(\frac{e_{i,j}}{(\Vert e_{i,j} - e^{\prime}_{i,j} \Vert_2)})
    \label{eqn:global_weights}
\end{equation}
where we update the relative co-occurrence using the previously observed co-occurrence frequency $e_{i,j}$ and the current co-occurrence frequency $e^{\prime}_{i,j}$ between k-mers $x_i$ and $x_j$. We bound the output of this co-occurrence formulation between -2 and 2 using a weighted update function $\sigma(\cdot)$. In our experiments, we set $\sigma(r) = 2 \sqrt{max(r - 1, 1)} + (min(r {-} 2, 2) {+} 2)$. The initial values of the co-occurrence are set to be $1$ when two k-mers $x_i$ and $x_j$ are first observed. If two k-mers never co-occur, then the frequency is $0$. 
Subsequent iterations of the update function, as illustrated in Algorithm~\ref{alg:cap}, increase the relative co-occurrence value to reinforce the presence of recurring k-mer pairs in the genome. 
Additionally, this formulation allows us to suppress spurious patterns that could be a product of noise while highlighting structures in the genome that are produced due to frequent co-occurrences of k-mers within the entire genome sequence.

\begin{table*}[t]
\caption{\textbf{Zero-shot Evaluation.} Accuracy is presented at different taxonomy levels for only \textit{unseen} species genomes. 
    TEPI-MGNet and TEPI-MG2V refer to genome TEPI trained with genome representations from MG-Net~\cite{aakur2021mg} and MG2V~\cite{aakur2021metagenome2vec}, respectively. TEPI-W2V refers to the use of \textit{word2vec} embeddings. 
    }
    \centering
    \begin{tabular}{|c|c|c|c|c|c|c|}
    \toprule
         \textbf{Hierarchy-level Acc.} & \textbf{top-K} & \textbf{TEPI-W2V} & \textbf{TEPI-MG2V} & \textbf{TEPI-MGNET} & \textbf{TEPI-WG} & \textbf{TEPI-Comp}\\
    \toprule
    \multirow{3}{*}{\textbf{Species}} & 1 & 3.90 & 3.66 & 10.39 & \textbf{11.69} & 7.79\\
 & 5 & 19.48 & 18.51 & 55.84 & 53.25 & \textbf{59.74}\\
 & 10 & 36.36 & 34.18 & 66.23 & 62.34 & \textbf{67.53}\\
 & 20 & 42.86 & 36.43 & 72.73 & \textbf{74.03} & 70.13\\
 \midrule
\multirow{3}{*}{\textbf{Genus}} & 1 & 44.16 & 37.09 & 46.75 & 48.05 & \textbf{49.35}\\
 & 5 & 53.25 & 42.60 & 59.74 & 58.44 & \textbf{64.94}\\
 & 10 & 59.74 & 54.36 & \textbf{68.83} & 63.64 & 67.53\\
 & 20 & 67.53 & 67.53 & 72.73 & \textbf{74.03} & 70.13\\
 \midrule
\multirow{3}{*}{\textbf{Family}} & 1 & 54.55 & 49.64 & \textbf{62.34} & \textbf{62.34} & 57.14\\
 & 5 & 64.94 & 61.69 & \textbf{72.73} & 68.83 & 70.13\\
 & 10 & 67.53 & 56.05 & 75.32 & 72.73 & \textbf{76.62}\\
 & 20 & 79.22 & 64.96 & 77.92 & 79.22 & \textbf{81.82}\\
 \midrule
\multirow{3}{*}{\textbf{Order}} & 1 & 59.74 & 54.36 & 70.13 & \textbf{71.43} & 63.64\\
 & 5 & 74.03 & 73.29 & \textbf{81.82} & 79.22 & 80.52\\
 & 10 & 77.92 & 69.35 & \textbf{85.71} & \textbf{85.71} & 83.12\\
 & 20 & 84.42 & 72.60 & 88.31 & \textbf{90.91} & 85.71\\
 
 \bottomrule
    \end{tabular}
    \label{tab:zsl_perf}
\end{table*}
We construct the pseudo-image $I_r$ by considering each pixel $p_{i,j}\in I_r$ to represent the relative co-occurrence of k-mers $x_i$ and $x_j$. Algorithm~\ref{alg:cap} provides a pseudocode for the pseudo-image construction process. This formulation allows us to represent the whole genome as an image and hence will enable us to capture long-range properties that have proven to be hard for sequence-based learning models such as LSTMs~\cite{hochreiter1997long} and Transformers~\cite{vaswani2017attention}. 
Additionally, it allows us to leverage the advances in computer vision, such as convolutional neural networks (CNNs)~\cite{He2016CVPR}, that can effectively capture the spatial relationships for effective representation learning. 
The resulting pseudo-image representation is given by 
\begin{equation}
    I_r(i,j) = \sum_{i=1}^{4^k}\sum_{j=i+s}^{4^k} 
        \begin{cases}
            255*f_r(i,j)/N, & \text{if } f_r(x_i,x_j) > \lambda_{m} \\
            0, & \text{otherwise}
        \end{cases}
    \label{eqn:image_pixels}
\end{equation}
where the pseudo-image is constructed by iterating over all possible k-mer combinations for a given window length $k$ and stride $s$. Each pixel $I_r(i,j)$ is assigned the output of $f_r(x_i, x_j)$ from Equation~\ref{eqn:global_weights}. 
The co-occurrences are normalized to be between $0$ and $1$ using $N$, which is the sum of all k-mer co-occurrences from all sequences in the genome; A cutoff parameter, $\lambda_{m}$, is introduced to remove any spurious patterns that can be introduced due to sequencing errors~\cite{laver2015assessing} which can flip base pairs randomly. We simulate an RGB image to leverage standard CNN architectures~\cite{He2016CVPR,simonyan2014very} for image processing by duplicating each pixel value to create an image of depth $3$. We set $\lambda_{m}=0$ in our experiments. 

The original MGNet~\cite{aakur2021mg} formulation was done at a sequence level for \textit{metagenomes}, where the pseudo-image was constructed for each sequence read. While useful in metagenomes (where each sequence can belong to a different species), this formulation does not consider global correlations between k-mers, essential in whole genome representations. 
For genome-level modeling, we explore three strategies for capturing the phylogenetic properties using pseudo-images. 
The first strategy, called \textit{TEPI-MGNet}, constructs a pseudo-image for each sequence read in the genome and then combines them by summing the pixel values individually. The resulting image is normalized by the sum of all genome-level k-mer co-occurrences, scaling each pixel value between 0 and 1.
The second strategy, known as \textit{TEPI-WG}, involves concatenating all sequence reads into a single, genome-level sequence. A pseudo-image is then constructed for each genome sample using this concatenated sequence.
The third strategy, named \textit{TEPI-Comp}, extends \textit{TEPI-MGNet} by considering both the sequence read and its reverse complement to capture the symmetry inherent in DNA structures. This strategy allows for additional cues in capturing phylogenetic variations at the genome level. Some examples of the pseudo-images generated using \textit{TEPI-Comp} are shown in Figure~\ref{fig:PI_Viz}, demonstrating the distinctive species-specific global patterns represented in the pseudo-images.
Empirically, in Section~\ref{sec:results}, it is found that incorporating the reverse complement improves the performance of \textit{TEPI-Comp} for unseen pathogens, with higher genome similarity of the predicted and actual species (Table~\ref{tab:genome_similarity}).

\begin{table*}[t]
    \caption{\textbf{Generalized zero-shot evaluation. 
    Accuracy is presented at different levels of the taxonomy. 
    }
    All the approaches are evaluated on genomes from both \textit{seen} and \textit{unseen} species for generalized genome recognition.}

    \centering
    \begin{tabular}{|c|c|c|c|c|c|c|}
    \toprule
         \textbf{Hierarchy-level Acc.} & \textbf{top-K} & \textbf{TEPI-W2V} & \textbf{TEPI-MG2V} & \textbf{TEPI-MGNET} & \textbf{TEPI-WG} & \textbf{TEPI-Comp}\\
    \toprule
    \multirow{4}{*}{\textbf{Species}} & 1 & 49.74 & 38.80 & 51.29 & 51.29 & \textbf{67.78}\\
& 5 & 72.94 & 68.56 & 90.21 & 86.86 & \textbf{92.01}\\
& 10 & 80.41 & 63.53 & 92.27 & 91.75 & \textbf{93.56}\\
& 20 & 84.54 & 73.55 & 94.07 & \textbf{94.59} & 94.07\\
  \midrule
\multirow{4}{*}{\textbf{Genus}} & 1 & 72.94 & 68.56 & 83.51 & 80.93 & \textbf{89.95}\\
& 5 & 86.60 & 84.87 & 90.98 & 87.89 & \textbf{93.04}\\
& 10 & 89.95 & 87.25 & 92.78 & 91.75 & \textbf{93.56}\\
& 20 & 92.53 & 90.68 & 94.33 & \textbf{94.59} & 94.07\\
  \midrule
\multirow{4}{*}{\textbf{Family}} & 1 & 78.61 & 63.67 & 89.43 & 86.34 & \textbf{91.49}\\
& 5 & 90.72 & 80.74 & \textbf{94.33} & 90.46 & {94.07}\\
& 10 & 92.78 & 74.23 & 94.85 & 94.07 & \textbf{95.36}\\
& 20 & 95.36 & 74.38 & 95.36 & 95.62 & \textbf{96.39}\\
  \midrule
\multirow{4}{*}{\textbf{Order}} & 1 & 81.19 & 65.76 & 91.24 & 88.40 & \textbf{92.78}\\
& 5 & 92.78 & 73.30 & \textbf{96.13} & 92.78 & \textbf{96.13}\\
& 10 & 95.36 & 87.73 & \textbf{96.91} & 96.65 & 96.65\\
& 20 & 96.91 & 76.56 & 97.42 & \textbf{97.94} & 97.16\\

 \bottomrule
    \end{tabular}
    \label{tab:gzsl_perf}
\end{table*}

\subsection{Genome-to-Taxonomy Mapping}~\label{sec:map}
Given the two representations, i.e., genome-level pseudo-images to capture phylogenetic variations at the nucleotide level and the taxonomy-aware, hierarchical embedding space, we need to find a mapping function $\phi: I(\mathcal{G}_i)\rightarrow\mathcal{E}_i$ that can map from the genome-level pseudo-image $I(\mathcal{G}_i)$ of each species $s_i\in \mathcal{S}$ to its corresponding taxonomic embedding $\mathcal{E}_i$, such that the genome representations possess the same compositional properties. 
While more complicated mechanisms, such as attention-based reasoning~\cite{aakur2021mg} and vision transformers~\cite{dosovitskiy2020image}, can be used, we use convolutional neural networks (CNNs)~\cite{simonyan2014very,He2016CVPR} to learn our mapping function $\phi(\cdot)$ for two primary reasons. First, CNNs have been shown to require significantly less training data to learn robust representations compared to vision transformers. Second, they are faster to train due to their inherent inductive biases to capture spatially relevant features. Empirically, we find CNNs to perform better in our ablation studies (Section~\ref{sec:ablation}). 
In this work, we use a 10-layer CNN network inspired by VGG-16~\cite{simonyan2014very}, where 5 blocks of 2 CNN layers each are interspersed with a max-pooling operation. A global average pooling layer is added at the end of the last convolutional block, followed by three fully connected layers to project down to the same dimension as the species-specific embedding. The network is trained to predict the taxonomy-aware embedding and capture the inter- and intra-class variations encoded within the taxonomy-aware embedding. We formulate the mapping function as a regression function from the pseudo-images $I(\mathcal{G}_i)$ to the embedding space $\mathcal{E}_i$. Hence, the training is an optimization for
\begin{equation}
  \mathcal{L} = \min_{\mathbf{W}_\phi} \vert\vert \mathbf{W}_\phi I(\mathcal{G}_i) - \mathcal{E}_i \vert\vert^2_d 
  \label{eqn:loss}
\end{equation}
where $\mathbf{W}_\phi$ is the set of learnable weights for the mapping function $\phi(\cdot)$ and $\mathcal{E}_i$ is a real-valued species-level $d$-dimension embedding. 
To aid in the computation, we add an L2 regularization layer to both the learned embedding $\mathcal{E}_i$ and the output of the mapping function $\phi(\cdot)$. This formulation, while simple, has several advantages. First, we offload the task of capturing phylogenetically relevant features to learn the embedding space. Hence, during training, we do not need to compute expensive mining of hard negative (or positive) pairs. Second, by formulating the mapping function as regressing to the hierarchical taxonomic embedding space, we can expand the vocabulary of the \textit{unseen} classes by adding the relevant embedding during inference. Finally, since the target embedding space is computed independently and fixed when training the mapping function, there is no need for additional supervision, such as in center loss~\cite{wen2016discriminative}.

\subsection{Implementation Details}
To learn the taxonomy-aware embedding, we use the $72,378$ species from the ``\textit{bacteria}'' subset from NCBI Taxonomy database~\cite{federhen2012ncbi} as our embedding search space to build our taxonomy tree to capture the compositional, phylogenetic relationships among all bacterial species. We generate 10 random walks for each species, with a maximum length of 100 visited nodes in each walk to generate the context. Having longer random walks allows us to capture the similarity in neighborhood structure for both highly related (such as from the same genus) and distantly related (i.e., from a different family) species. 
We set the initial learning rate to be $4\times 10^{-5}$, and we first perform a \textit{cold start}, i.e., for five epochs, the learning rate is set to be $4\times 10^{-12}$ and then increased. We use a k-mer of size 6 and stride of 10 to build the pseudo-images and resize the image to $512\times 512$. 
All hyperparameters are found using a grid search and kept constant for all experiments. 
All networks are trained for 500 epochs with early stopping. 
\section{Experimental Evaluation}\label{sec:results}
In this section, we present the results from the experimental evaluation of the proposed approach. We begin with a description of the data, followed by a discussion on the various evaluation settings and considered baselines for comparison. We follow with the presentation of the quantitative and qualitative results in various evaluation settings.

\subsection{Data Collection}
\textbf{Data Collection.} For our evaluation, we primarily focus on the bacterial genomes associated with common veterinary diseases, whose recognition is vital in diagnosing diseases that can yield severe economic distress. Table~\ref{tab:data} summarizes the dataset statistics. Specifically, we choose $93$ species, split into seen ($58$ species) and unseen ($35$ species). 
We downloaded \textit{complete} genomes for each species from the NCBI RefSeq~\cite{o2016reference} database, a publicly available database of annotated genomic sequences. Each genome is annotated using NCBI's prokaryotic genome annotation pipeline~\cite{tatusova2016ncbi}. We obtained a maximum of $10$ genomes per species from the RefSeq database and split them into train and test sets. If a species has at least ten genomes, five were chosen for evaluation as part of the \textit{seen} group of species, and the remaining five were used for training. Species with less than ten genomes were used as part of the \textit{unseen} classes for testing. To ensure proper coverage, we considered species with different phylogenetic characteristics such as gram stain (positive vs. negative), shape (Cocci vs. Bacilli vs. Spirilla), and oxygen dependency (aerobic vs. anaerobic), among others. 

\begin{table}[t]
    \caption{\textbf{Dataset Statistics}. We cover a wide range of seen and unseen genomes with coverage at different levels of the hierarchy. }
    \centering
    \resizebox{\columnwidth}{!}{
    \begin{tabular}{|c|c|c|c|c|c|c|}
         \toprule
         \textbf{Split} & \textbf{Class} & \textbf{Order} & \textbf{Family} & \textbf{Genus} & \textbf{Species} & \textbf{\# Train/Test}\\
         \toprule
         Seen & 15 & 25 & 36 & 52 & 58 & 290/290\\
         Unseen & 11 & 16 & 17 & 19 & 35 & -/143\\
         \bottomrule
    \end{tabular}
    }
    \label{tab:data}
\end{table}
\begin{figure*}[t]
    \centering
         \includegraphics[width=0.8\textwidth]{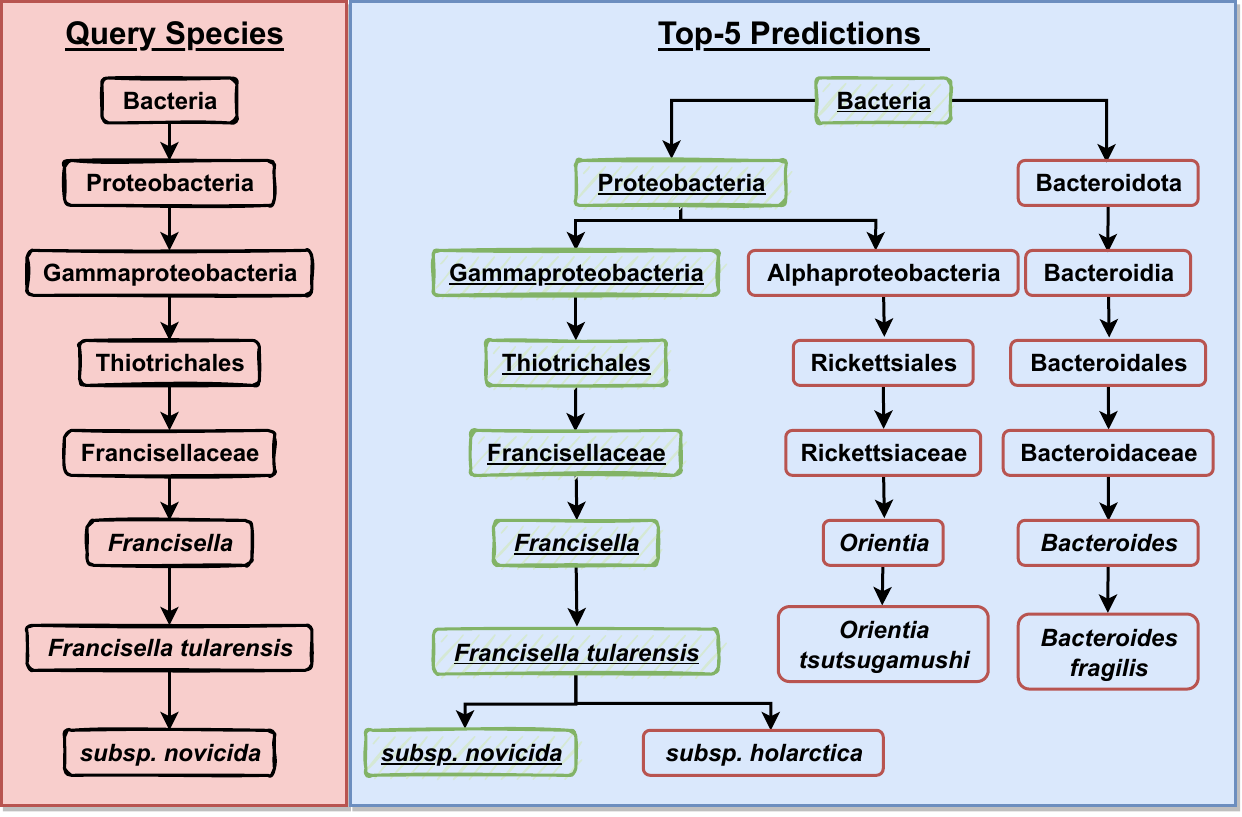}  \\
    \caption{\textbf{Qualitative visualization} of the predicted taxonomy tree from top-5 predictions (on the right) for a given query genome (left) for an unseen \textit{sub-species}. The top-1 prediction is underlined, while the correct predictions are in green. Note: \textit{Francisella tularensis} was also part of the top-10 predictions, along with the two sub-species, while \textit{subsp. novicida} was \textit{unseen} during training.}
    \label{fig:qual_subsp}
\end{figure*}

\subsection{Evaluation Settings and Baseline Approaches}
\textbf{Evaluation settings.} To evaluate the generalization capabilities of the approach, we adopt two settings following prior zero-shot learning works~\cite{merrillees2021stratified,wu2020multi}. In the first setting, i.e., \textit{zero-shot} evaluation, the approaches were trained on a set of \textit{seen} classes whose reference genomes and labels are available. During the evaluation phase, the frameworks are presented with genomes from \textit{unseen} classes, i.e., the genomes are not part of the training process. The goal is to classify the genome into one of the unseen species. In the \textit{generalized} zero-shot setting, genomes from \textit{both} seen and unseen species are presented, and the goal is to recognize the species corresponding to each genome correctly. The accuracy for top-K predictions is used to quantify the performance. Due to the large search space and limited labeled data, we evaluate at $K=\{1, 5, 10\}$. Additionally, for a fair comparison, we use a standard k-Nearest neighbor classifier from Sci-Kit Learn~\cite{pedregosa2011scikit}, trained on the embedding space with a neighborhood of 20.

\textbf{Baselines.} Since this is the first attempt at zero-shot genome classification, we propose a variety of baselines to compare against the proposed framework. First, in \textit{TEPI-W2V}, we train a \textit{word2vec} model~\cite{mikolov2013distributed} on simple chains of taxonomic lineages for each species to obtain the latent embedding space and use the proposed genome-wide pseudo-imaging as the genome representation. We pursue state-of-the-art deep learning models based on transformers~\cite{vaswani2017attention} as proposed in MG2V~\cite{aakur2021metagenome2vec} for learning representations from genome sequences. Finally, we use different variations of genome-wide pseudo-imaging proposed in Section~\ref{sec:PI}. Considered together, these baselines provide a first look at how deep learning models can be used to tackle zero-shot genome recognition. 

\subsection{Quantitative Evaluation}\label{sec:quant}
\textbf{Zero-shot Evaluation.} 
We first present the results of the zero-shot evaluation in Table~\ref{tab:zsl_perf}. We evaluate the ability of the approaches to recognize the query genome at different levels of the hierarchy, i.e., at the species, genus, family, and order raGENOMEnks of the taxonomy. While we are more interested in the species level accuracy, the accuracy at the other levels provides a measure of how well the proposed embedding space (Section~\ref{sec:embed}) captures the taxonomic hierarchy. As can be seen, the approaches with the proposed TEPI embedding space significantly outperform the \textit{word2vec} baseline (\textit{TEPI-W2V)} at all accuracy metrics, indicating that na\"ive taxonomic embeddings do not generalize well to unseen species. Similarly, other genome representations, such as MG2V, perform poorly while performing better than the \textit{word2vec} model, indicating that the local contexts from the sequence level do not scale to genome-wide representations. Interestingly, the complement-aware pseudo-imaging has poor top-1 accuracy at the species level but improves dramatically at the genus and family levels. We also find that although the performance of all approaches is similar at the higher levels of the hierarchy, such as order and family, \textit{TEPI-MGNET} with the composite MGNET representations does better. Beyond the order level of the taxonomy, i.e., at the class, phyla, and kingdom levels, the performance of all TEPI-based models reaches $100\%$ at $K{=}5$, indicating that the learned embedding captures the compositional relationship between the species. 

Note that MG2V and MG-NET are supervised learning frameworks that take several hours (almost 20 for MG2V and 10 for MG-NET) of pre-training and 4 hours of fine-tuning to learn robust features and do not generalize to unseen genomes since the predictions are limited to those seen during training. Our modification to MG2V, TEPI-MG2V, allows us to scale the framework to unseen genomes, although it does struggle with learning long-range dependencies found in whole genome sequences. Using metadata such as structural features used in MG2V and MG-NET could potentially improve the performance of TEPI. Similarly, integrating the taxonomy-aware embedding into supervised learning using semantic codes or knowledge-infused learning techniques can help machine learning-based models generalize beyond a limited vocabulary. 
Section~\ref{sec:qual} discusses some interesting success and failure cases of TEPI and comparative frameworks. 

\begin{table}[t]
    \centering
    \caption{\textbf{Whole Genome similarity} based on BLAST~\cite{ye2006blast} for the top-1 predicted species for \textit{seen} and \textit{unseen} query genomes. * indicates  ($p{>}0.05$)}
    \begin{tabular}{|c|c|c|c|c|}
         \toprule
         \multirow{2}{*}{\textbf{Approach}} & \multicolumn{3}{|c|}{\textbf{Genome Similarity}} & \textbf{Spearman}\\
         \cline{2-4}
         & \textbf{Seen} & \textbf{Unseen} & \textbf{Overall} & \textbf{Coefficient}\\
         \toprule
        TEPI-W2V & 77.93 & 78.51 & 78.04 & 14.25\\
        TEPI-MG2V & 79.91 & 79.31 & 79.79 & \underline{10.26}*\\
        TEPI-MGNET & 97.18 & \textbf{87.58} & 95.27 & \textbf{77.55}\\
        TEPI-WG & 97.92 & 87.43 & 95.89 & 73.95\\
        TEPI-Comp & \textbf{99.59} & 86.97 & \textbf{97.04} & 57.47\\
         \bottomrule
    \end{tabular}
    \label{tab:genome_similarity}
\end{table}

\textbf{Generalized Zero-shot Evaluation}
Table~\ref{tab:gzsl_perf} summarizes the performance of the considered baselines in the generalized zero-shot learning setting, where the query genome at test time can come from \textit{both} seen and unseen species. It is interesting to note that \textit{TEPI-Comp}, the complement-aware genome representation trained with the proposed taxonomy-aware embedding space, significantly outperforms all baselines. We attribute it to the fact that it obtains $82.7\%$ top-1 accuracy at the species level and $100\%$ top-1 accuracy at all other taxonomy levels for \textit{seen} classes, even when training with only $5$ reference genomes per class. \textit{TEPI-WG} and \textit{TEPI-MGNET} achieve $61.5\%$ and $61.5\%$ top-1 accuracy at the species level and an average of $90.7\%$ and $95.1\%$ top-1 accuracy at other levels, respectively. We hypothesize that the reverse complement representation carries distinguishing characteristics that allow easier translation from the genome to the taxonomy-aware embedding space for genomes from seen classes. We find that the \textit{word2vec} embedding space works better in the generalized zero-shot setting than the MG2V representations but does not outperform the pseudo-image-based approaches. We again observe the performance saturation for all approaches at the family and order levels for top-5 and top-10 predictions, indicating that the taxonomy is preserved in the embedding space. 

\textbf{Improved Latency.} We also compared the computational efficiency of using the proposed TEPI framework for zero-shot genome recognition. B. We observe that, on average, the time taken to compare the query genome with other reference genomes is lower for TEPI than BLAST. This difference is more apparent as the number of reference genomes for comparison increases. In TEPI, the significant cost is the construction of the pseudo-image, which takes 3 seconds per genome. The cost for extracting features is 300 ms on an NVIDIA RTX 3060 GPU, and the kNN-based search takes approximately 800 ms using Sci-Kit Learn for searching over 1000 candidate genomes. The total execution time per query is, on average, 8 seconds, averaged over 100 runs. For BLAST, the gold standard bioinformatics platform, searching over 1000 candidate genomes takes around 25 minutes (averaged over 100 runs) on a 32-core AMD ThreadRipper CPU. Similarly, the closely related MG2V, titled TEPI-MG2V, takes 10 seconds to extract features per query and around 5 minutes to search over 1000 candidate genomes. 

\subsection{Qualitative Analysis.}\label{sec:qual} 
In addition to the quantitative analysis, we present some qualitative results to dissect the performance of the proposed TEPI approach. Specifically, we present two cases - (i) genome classification at the \textit{unseen sub-species} level and (ii) classification of a genome from an \textit{unseen genus} and species. Figure~\ref{fig:qual_subsp} presents a query genome from \textit{Francisella tularensis subsp. novicida}, whose taxonomy is shown on the left. The consolidated taxonomy tree from the top-5 predictions made by \textit{TEPI} is shown on the right. The correct and incorrect classifications at each level are highlighted in green and red, respectively, whereas the top-1 prediction is underlined. Although the subspecies \textit{novicida} is unseen, the approach was able to retrieve very relevant neighbors (top 3 predictions were \textit{Francisella tularensis subsp. novicida}, \textit{Francisella tularensis} and \textit{Francisella tularensis subsp. holarctica}) while placing the correct subspecies at the top-1 prediction. This performance is remarkable considering that the other subspecies \textit{holarctica} was part of the training set yet was not the top prediction. This result indicates that subtle variations at both genome and taxonomy levels are captured in TEPI. Similarly, in Figure~\ref{fig:qual_genus}, TEPI was able to retrieve highly relevant species to a genome query from an unseen \textit{genus} (\textit{Biberstinia)}. Interestingly, all top-5 predictions were placed within the same family \textit{Pasteurellaceae} of the target species, with the correct prediction at the third position on the prediction order.

\textbf{Genome Similarity.} To qualitatively analyze the predictions of the approach, we use BLAST~\cite{ye2006blast} to align and compute whole genome similarity between the predicted and actual species during inference. Table~\ref{tab:genome_similarity} summarizes the results. It can be seen that the genome similarity between the predictions of the pseudo-image-based approaches is significantly better than that from MG2V. Similarly, the \textit{word2vec}-based model performs considerably worse than the proposed taxonomy-aware embedding. TEPI-Comp's predictions have significantly higher genome similarity for \textit{seen} classes, whereas, for \textit{unseen}, TEPI-MGNET performs better. From these results, we can see that even though the top-1 performance of the TEPI-based approaches can be lower at the species level ($11.69$ for TEPI-WG), the genome similarity of the top-1 is more than $87\%$, which indicates that the pseudo-imaging and taxonomy-based embedding functions capture the underlying relationships from the genome sequences, which is our primary focus. 

We also evaluate whether the distance between the predicted embeddings for two genomes can serve as a proxy for BLAST-based similarity measures, which can be costly due to expensive alignment steps. We take the dot product between the embedding of the predicted label and the actual label under the generalized zero-shot setting and compute Spearman's rank correlation coefficient to ascertain its correlation with BLAST-based genome similarity scores. A higher correlation indicates better alignment between actual genome similarity and the model's notion of similarity. As seen from Table~\ref{tab:genome_similarity}, TEPI-MGNET has the highest correlation, followed by TEPI-WG. Interestingly, TEPI-Comp, which has the highest performance, has a poor correlation with the BLAST-based similarity scores. All numbers reported are statistically significant ($p{<}0.05$), except TEPI-MG2V, which has poor performance but is not statistically significant ($p{=}0.219$). The \textit{word2vec}-based embedding model (TEPI-W2V) does poorly on this as well, indicating that it does not capture the genome-level similarity effectively in its embedding function. Note that this distance is computed based on the mapping function, defined in Section~\ref{sec:map}, and not on the pseudo-image itself. In ablation studies (Section~\ref{sec:ablation}), we see that using a supervised contrastive learning loss~\cite{khosla2020supervised} in addition to the loss defined in Equation~\ref{eqn:loss} without hard negative mining improves the genome similarity of predicted classes. Contrastive learning~\cite{wang2022negative} with a nuanced negative mining scheme could potentially improve this performance.
Additionally, exploiting the structural properties encoded in the pseudo-image using self-supervised representation learning~\cite{bengio2013representation} could provide a better correlation with genome similarity.  

\subsection{Ablation Studies.}\label{sec:ablation} 
\begin{table}[t]
    \caption{\textbf{Ablation studies} to evaluate the effect of genome encoder, embedding functions, and using a contrastive learning loss. }
    \label{tab:ablation}
    \centering
\resizebox{0.99\columnwidth}{!}{    
    \begin{tabular}{|c|c|c|c|c|c|}
    \toprule
    \textbf{Genome} & \textbf{Taxonomy} & \textbf{Contrastive} & \multicolumn{2}{|c|}{\textbf{Species}} & \textbf{Unseen Genome} \\
    \cline{4-5}
    \textbf{Encoder} & \textbf{Embedding} & \textbf{Learning?} & \textbf{Unseen} & \textbf{Seen} & \textbf{Similarity}\\
    
    \toprule
    MG-Net~\cite{aakur2021mg} & TEPI-Comp & \ding{55} & 7.79 & 61.42 & 86.97 \\
    MG-Net~\cite{aakur2021mg} & TEPI-MGNET & \ding{55} & 10.39 & 51.29 & 87.58 \\\midrule
    MG-Net~\cite{aakur2021mg} & TEPI-Comp & \ding{51} & 6.09 & 83.62 & 86.67 \\
    MG-Net~\cite{aakur2021mg} & TEPI-MGNET & \ding{51} & 5.19 & 62.38 & 87.05 \\
    \midrule
    ResNet~\cite{He2016CVPR} & TEPI-Comp & \ding{55} & 8.09 & 62.01 & 87.09 \\
    ViT~\cite{dosovitskiy2020image} & TEPI-Comp &  \ding{55} & 7.93 & 83.75 & 86.99 \\
         \bottomrule
    \end{tabular}
    }
\end{table}
In addition to quantitative and qualitative evaluation, we systematically analyze the effect of TEPI's components on its performance. 
Table~\ref{tab:ablation} summarizes the results. 
First, we evaluate the impact of using different visual encoder architectures for genome-to-taxonomy mapping (Section~\ref{sec:map}). In our primary experiments (Section~\ref{sec:quant}), we used the VGG-style encoder proposed in MG-Net~\cite{aakur2021mg} for fair comparison with prior approaches. 
However, research in the computer vision community has resulted in more nuanced architectures such as the ResNet~\cite{He2016CVPR}, a convolutional neural network that has been successfully scaled to a larger, more efficient architecture, and the vision transformer (ViT)~\cite{dosovitskiy2020image}, an alternative to CNN-based architectures that uses transformers~\cite{vaswani2017attention} to learn robust features from images. Using ResNet and ViT results in marginal improvements over the simpler VGG-style encoders. 
We attribute this improvement to the features learned from the large-scale pre-training on natural image data. However, we do not observe considerable improvements, possibly due to the out-of-domain nature of pseudo-imaging compared to the natural images found in ImageNet. Large-scale pre-training can help improve the performance of larger vision encoders on genome-based pseudo-images. 

In addition to the mapping loss proposed in Equation~\ref{eqn:loss}, we experiment with adding a supervised contrastive learning~\cite{khosla2020supervised} loss to encourage the network to learn more discriminative features for each species. This loss function is a form of metric learning~\cite{musgrave2020metric} where the goal is to reorganize the latent space by forcing the representations from similar inputs together and pushing inputs from other classes away. We see that the addition of this contrastive learning loss significantly affects TEPI's performance on species \textit{seen} during training, providing gains of almost $20\%$ on TEPI-Comp and $10\%$ on TEPI-MGNET. However, the performance on \textit{unseen} classes, i.e., the generalization to novel species, takes a severe hit ($5\%$ on unseen classes and $2\%$ on seen classes). 
This effect could be due to the fact that supervised contrastive loss uses multiple positive samples and does not use hard negatives to reorganize the latent space. The performance could possibly be improved with a more nuanced selection of negative examples for contrastive learning by leveraging the highly complex similarity measures across species. 

\subsection{Limitations and Opportunities}\label{sec:discussion} 
While the proposed TEPI framework does well on zero-shot and generalized zero-shot recognition of whole genomes (Section~\ref{sec:quant}) and provides interesting insights on capturing the genome similarity (Section~\ref{sec:qual}) using pseudo-imaging, we observe that there are some limitations that can be the source of future work in zero-shot genome classification. First, our experiments with contrastive learning (Section~\ref{sec:ablation}) yielded interesting results that emerge when adding a metric learning loss to TEPI. While the performance did not improve much on unseen genomes, the nature of failures provides critical insights into how the framework could benefit from metric learning. Specifically, we observe that the accuracy at the genus level is improved on unseen genomes, indicating that the clustering occurs at higher levels of the taxonomy. Hence, introducing a hard negative sampling at the fine-grained level can lead to better performance. Second, the pseudo-image formalism offers a new way of representing genomes but suffers from the limitation that the image captures patterns at a single stride length. Extending pseudo-imaging to handle multiple stride lengths could lead to more robust features that can yield better performance on taxonomic profiling. Finally, TEPI is currently designed to work with whole genomes, whereas critical diagnostic applications could benefit from performance on 16S and 23S gene sequencing reads and raw \textit{metagenome} sequences. Whole genome sequences generally possess longer, more extensive sequences than 16S/23S reads and metagenome sequences, yielding richer context in pseudo-imaging. Metagenome sequences and 16S/23S sequences have shorter context windows that can require re-formulation of the pseudo-image representation to enable its functionality. 

TEPI's robust representation learning paradigm and generalization capabilities offer several opportunities for integration into bioinformatics pipelines such as BLAST. For example, we can envision TEPI as an initial, coarse filter to reduce BLAST's search space and overall latency in taxonomic profiling. Section~\ref{sec:qual} shows that the computational overhead is significantly lower for TEPI compared to BLAST and other machine learning models such as MG2V. Given that top-10 \textit{genus-level} prediction for seen classes is more than $90\%$, using TEPI as a filter to narrow down BLAST's search space could considerably reduce the latency for taxonomic profiling while improving performance. 
While we only explored the taxonomic classification of whole genome sequences, we envision TEPI as a learning paradigm for capturing semantically relevant features for downstream tasks such as disease and pathogen characterization tasks such as predicting potential drug resistance, pathogenic potential, and disease progression. For example, disease characterization models such as Metagenome2Vec~\cite{queyrel2021towards,nguyen2019metagenome} could benefit from the recognition of pathogens present in the genome sequences and result in more nuanced prediction of disease outcomes from raw genome data. Our future works are focused on capturing the genome-level similarities between species for more robust recognition and zero-shot transfer. 
\section{Conclusion and Future Work}\label{sec:conclusion}
In this work, we presented TEPI, one of the first works to address the problem of zero-shot genome recognition from extremely scarce data. Using a hierarchical graph structure extracted from the species taxonomy, the proposed embedding space provides a compositional representation that can efficiently be used for zero-shot recognition of whole genomes. An image-based representation of the whole genome captures subtle intra- and inter-species variations that can be translated to the learned embedding space in a simple yet highly effective framework with only five labeled examples per species. We aim to extend TEPI to handle zero-shot recognition of 16S and 23S rRNA gene sequencing, which are easier to sequence and more readily available compared to whole genome sequences, to drive the development of the next generation of point-of-care diagnostic platforms. 

\section*{Acknowledgements.} 
This research was supported in part by the US Department of Agriculture (USDA) grants AP20VSD and B000C011.
We also thank Dr. Arunkumar Bagavathi, Dr. Sai Narayanan, and Mr. Udhav Ramachandran for thoughtful discussion on pseudo-imaging for zero-shot taxonomic profiling. 
\bibliography{egbib.bib}
\bibliographystyle{IEEEtran}
\begin{IEEEbiographynophoto}{Sathyanarayanan N. Aakur} is an Assistant Professor at the Department of Computer Science and Software Engineering at Auburn University. He received his B.Eng. degree in Electronics and Communication Engineering from Anna University, Chennai, India, in 2013. He received an M.S. in Management Information Systems and a Ph.D. in Computer Science and Engineering from the University of South Florida, Tampa, in 2015 and 2019, respectively. He received the prestigious National Science Foundation CAREER award in 2022 and has been Associate Editor for IEEE Robotics and Automation Letters since 2021. His research interests lie in neuro-symbolic reasoning for open-world visual understanding and deep learning applications for computational biology. 
\end{IEEEbiographynophoto}

\begin{IEEEbiographynophoto}{Vishalini Ramnath} is the Assistant Director of Undergraduate Programs at the Department of Computer Science and Software Engineering at Auburn University. She received her B.Eng. degree in Electronics and Communication Engineering from Anna University, Chennai, India 2013. She received an M.S. in Computer Science from the University of Illinois, Chicago, and a Ph.D. in Computer Science and Engineering from the University of South Florida, Tampa, in 2016 and 2020, respectively. Her research interests include machine learning for the Internet of Things and deep learning applications for genomics.
\end{IEEEbiographynophoto}

\begin{IEEEbiographynophoto}{Priyadharsini Ramamurthy} is currently pursuing her Ph.D. degree in Computer Science from Oklahoma State University. She received a B.Eng. degree in Computer Science from Anna University, Chennai, India, in 2006. She received an M.S. degree in Engineering Management from International Technological University. Her research interests include Natural Language Understanding and deep learning applications.
\end{IEEEbiographynophoto}

\begin{IEEEbiographynophoto}{Akhilesh Ramachandran} is an Associate Professor with the Department of Veterinary Pathobiology at Oklahoma State University.  He also serves as the Head of the Microbiology and Molecular Diagnostics Sections at the Oklahoma Animal Disease Diagnostic Laboratory. He received the B.V.Sc. \& A.H. degree from Kerala Agricultural University, Kerala, India, in 1998. He received a Ph.D. in Veterinary Biomedical Sciences in 2003 from Oklahoma State University. His research interests include the development of novel molecular diagnostic protocols and platforms for infectious disease diagnosis.  
\end{IEEEbiographynophoto}
\end{document}